**ARTICLE**

# A Robust Governance for the AI Act: AI Office, AI Board, Scientific Panel, and National Authorities

Claudio Novelli[1] , Philipp Hacker[2], Jessica Morley[3], Jarle Trondal[4,5,6] and Luciano Floridi[3,1]

[1]Department of Legal Studies, University of Bologna, Bologna, IT, USA, [2]European New School of Digital Studies, European University Viadrina, Frankfurt (Oder), Germany, [3]Digital Ethics Center, Yale University, New Haven, CT, USA, [4]Department of Political Science and Management, University of Agder, Kristiansand, Norway, [5]ARENA Centre for European Studies, University of Oslo, Oslo, Norway and [6]Institute of European Studies, University of California, Berkeley, CA, USA

**Corresponding author:** Claudio Novelli; Email: claudio.novelli@unibo.it

**Abstract**

Regulation is nothing without enforcement. This particularly holds for the dynamic field of emerging technologies. Hence, this article has two ambitions. First, it explains how the EU's new Artificial Intelligence Act (AIA) may be implemented and enforced by various institutional bodies, thus clarifying the governance framework of the AIA. Second, it proposes a normative governance model, providing recommendations to ensure uniform and coordinated execution of the AIA and the fulfillment of the legislation. The article explores how the AIA may be implemented by national and EU institutional bodies, encompassing longstanding bodies, such as the European Commission, and those newly established under the AIA, such as the AI Office. It investigates their roles across supranational and national levels, emphasising how EU regulations influence institutional structures and operations. These regulations may not only *directly* dictate the structural design of institutions but also *indirectly* request administrative capacities needed to enforce the AIA.

## I. Introduction

The effective implementation of the Artificial Intelligence Act (AIA) throughout the European Union (EU) depends on a uniform, coordinated, and well-funded governance setting.[1] This is especially important given the increasing need for harmonised regulatory application in the digital sector, as emphasised by EU policymakers due to the numerous laws already enacted.[2] For this purpose, the AIA, notably in Chapter VII ('Governance'), underscores the role of different institutional bodies, supranational and national, such as the AI Office, the European AI Board, the Advisory Forum, the Scientific Panel, and (two) national competent authorities in each Member State. Close coordination between these bodies is crucial for implementing and enforcing the AIA's rules across all Member

---

CN's contributions were supported by funding provided by Intesa Sanpaolo to the University of Bologna.

[1] This is explicitly stated by the AIA at Recital 148.

[2] J Tar, "Implementing Tech Rulebooks Should Be Top Digital Policy Priority, EU Countries Urge" <https://www.euractiv.com/section/digital/news/implementing-tech-rulebooks-should-be-top-digital-policy-priority-eu-countries-urge/> (accessed 25 March 2024).

States. This interaction should also guarantee compatibility with other EU regulations to avoid redundancy and antinomies.

This article explores how the AIA may be implemented by the EU institutional bodies, encompassing longstanding bodies, such as the European Commission (Commission), and those newly established under the AIA, such as the AI Office. It investigates their roles across supranational and national levels, emphasising how EU regulations influence institutional structures and operations. These regulations may not only *directly* dictate the structural design of institutions but also *indirectly* request administrative capacities needed to enforce the AIA.[3]

These deliberations share an important dynamic aspect: bodies enforcing the AIA will oversee activities in various sectors due to the rapidly expanding reach of AI into all products and services. Interconnections with the enforcement of other recent EU legislations and the digital industry, such as the Digital Services Act (DSA), are bound to arise. Hence, at the EU and the national level, AIA enforcement bodies, such as the AI Office and specific national regulators, may ultimately be considered the nucleus of more encompassing "digital agencies," bundling competencies and expertise across various digital instruments. This raises the stakes of designing these entities wisely.

Despite existing research on how EU regulatory governance influences national governance processes, we know little about how EU policy regulations shape states' enforcement infrastructures – that is, the organisational design of public administration.[4] To explore this, we will delve into the institutional design of these bodies, which includes the structure, competence, (division of) tasks, funding, and allocation of responsibilities.

The normative framework is becoming more established, especially after the consolidation of the AIA, but there remains scope for additional adjustments in the phase of implementing and delegated acts. The implementation stage enables the Commission and, on rare occasions, the Council of the European Union to fine-tune non-essential aspects of the legislation. Experts nominated by each Member State are consulted before adopting these acts.

Against this background, this article has two ambitions. First, it explains how the AIA may be implemented and enforced by supranational and national bodies, thus illuminating the governance framework of the AIA. Second, it proposes a normative governance model, providing recommendations to ensure uniform and coordinated execution of the AIA and the fulfillment of the legislation. The awareness informs these recommendations of the uncertainties surrounding the future development of AI technologies and their social impacts. This perspective leads us to endorse a model of governance characterised by its robustness. Robustness, in this sense, implies the ability to uphold core functions, purposes, and values and maintain critical structural or operational architectures in the face of disruptive perturbations through adaptation and innovation.[5]

---

[3] Government structuring is a national prerogative, coined as national "administrative sovereignty" in extant literature. The latter is understood as the legal right of final decision on the structuring of government and the "assertion of control over recognisable administrative mechanisms of a government separate from the comprehensive operation of a nation" in K Muth, "The Potential and Limits of Administrative Sovereignty", *The Oxford Handbook of Global Policy and Transnational Administration* (Oxford Academic 2019) 60 <https://academic.oup.com/edited-volume/28088/chapter/212140358> (accessed 17 April 2024).

[4] A Benz, J Broschek and M Lederer (eds), *A Research Agenda for Multilevel Governance* (Elgar 2021) <https://www.e-elgar.com/shop/gbp/a-research-agenda-for-multilevel-governance-9781789908367.html> (accessed 17 April 2024); M Egeberg and J Trondal, "Researching European Union Agencies: What Have We Learnt (and Where Do We Go from Here)?" (2017) 55 JCMS: Journal of Common Market Studies 675; Muth (n 4).

[5] C Ansell and others, "Robust Governance in Turbulent Times" (2024) Elements in Public Policy <https://www.cambridge.org/core/elements/robust-governance-in-turbulent-times/AB44DBE9AA636390EC114E8A428BF188> (accessed 21 March 2024).

The article is structured as follows. Section II discusses general considerations for the design of agencies and bodies tasked with AI legislative enforcement and supervision. Section III reviews the critical components for implementing the AIA, focusing on the Commission's implementing and delegated acts. Section IV examines the supranational entities overseeing the AIA, including the AI Office, the AI Board, the Advisory Forum, and the Scientific Panel, proposing measures to streamline the governance framework to eliminate redundancies. Section V analyses national authorities' roles – highlighting notifying authorities, notified bodies, and market surveillance authorities. Section VI offers a set of recommendations derived from the analysis performed. The conclusion is presented in Section VII.

## II. General considerations: designing robust governance for the AIA

This section discusses the potential goals, structures, interdependencies, and challenges of establishing a multilevel governance framework for AI in the EU and Member States.

### II.1. EU level

At least three institutional design options are available at the EU level to establish executive capacities for regulating and enforcing AI.

Option 1 suggests a *centralised* institutional design to incorporate tasks related to AI regulations within the remit of the Commission – notably within its departments, i.e., its Directorates-General (DG). Such an approach could involve establishing a new DG (or a new unit within it) or reforming an existing one by increasing its policy portfolio to incorporate AI (e.g., Connect A responsible for Artificial Intelligence and Digital Industry). Implementing this structure would enhance the Commission's ultimate control, oversight, and management of AI policy regulation and enforcement activities.

Option 2 is a *decentralised* institutional design incorporating AI-related tasks in EU-level agencies. Similar to the Commission, this could involve either establishing a new AI agency at arm's length from the Commission or a reformed EU agency incorporating AI tasks in its task portfolio. Option 2 would thus leave the Commission with less control, oversight, and day-to-day management.

Option 3 implements a *hybrid* institutional design with AI-related tasks established within the Commission in a designated DG, with one or several EU-level agencies governing at arm's length from the Commission DG and working closely with other relevant DGs. Existing literature suggests that most decentralised EU-level agencies keep strong ties to what they consider their corresponding or "parent" DG.[6] We consider this the option most likely to result in suitably robust AI governance as it balances the strengths and weaknesses of the other two options.

### II.2. Member state level

Establishing national agencies responsible for enforcing AI regulations, two of which the AIA has already introduced, presents three institutional design options for consideration.

Option 1 would establish a new national agency dedicated to AI regulation enforcement. Its main benefit is creating a centralised body designed explicitly for AI oversight, attracting personnel with skills tailored to AI's distinct requirements. However, it may lack industry-specific expertise and risk detachment from the intricacies of different sectors.

---

[6] Egeberg and Trondal (n 5).

Moreover, the urgency of the AIA's application, with the first four editions effective from the end of 2024, makes the typically lengthy process of legally and institutionally establishing a new agency a significant drawback.

Option 2 would simply assign the AIA's enforcement to an existing agency. It capitalises on the existing organisational framework and sectoral knowledge. For instance, the banking sector has utilised machine learning models for decades,[7] and banking authorities have significant experience in testing and supervising these models, at least since the 2008 financial crisis and the accompanying overhaul of the EU financial services and banking regulation.[8] However, this could lead to disputes over mandate allocation and potentially narrow the focus to specific sectors, ignoring the AI Act's broad applicability.

Option 3 would merge centralised expertise with sectoral insights by establishing a new "competence center"[9] within an existing authority with AI experience, such as a banking or network regulator. The competence center would temporarily or permanently bring together AI experts from different backgrounds to form interdisciplinary teams (e.g., legal experts and computer scientists) on specific cases. This approach aims to integrate comprehensive AI knowledge with in-depth sectoral understanding despite potential recruitment challenges, particularly for technical positions.

As the implementation of the AIA at the State Member level is ongoing, the choice among the proposed options remains uncertain. Nonetheless, it is possible to speculate on the effectiveness of these options. Using Germany as an illustrative example, Options 2 and 3 are more robust than Option 1. Political dynamics and the convenience of existing frameworks may lead decision-makers to favor Option 2. Key agencies considered for AIA oversight include the Federal Office for Information Security (BSI), the Federal Network Agency (BNetzA), and state data protection authorities, each with specific strengths and challenges. BSI's technical expertise is valuable for identifying AI risks, though it may not cover all regulatory aspects. BNetzA has a broad regulatory scope but could lack AI-specific expertise. State data protection authorities are well-versed in privacy issues but might not fully address AI's broader impacts. However, Option 3 promises a more balanced mix of agility, specialised knowledge, and sector-wide understanding. Thus, an Option 3 competence center, linked to the banking regulator (BaFin) or one of the other agencies mentioned (e.g., BNetzA), could offer the necessary flexibility and sectoral insight for effective AI regulation, leveraging BaFin's experience in managing machine learning within financial oversight to meet the AI Act's requirements, or BNetzA's expertise in governing infrastructure and platforms (as the new national Digital Services Coordinator enforcing the Digital Services Act).

The effectiveness of these institutional designs may depend on the evolving framework at the EU level, which may ultimately determine what constitutes a robust institutional design at the national level.

### II.3. Multilevel: relationship between EU and national levels

Multilevel administrative systems consist of relatively stable arrangements of bureaucratic institutions and processes that span levels of government. Yet, depending on the chosen institutional designs, different multilevel governing relationships are likely to unfold across levels of governance. Extant literature suggests multilevel governance processes are particularly affected and biased by two institutional conditions. One is the degree of *administrative decentralisation* – e.g., 'agencification'[10] – of national-level government structures: the more task portfolios are hived off from ministries to agencies at the national level, the more likely it is that these agencies, in turn, establish governing relationships with 'their' sister agencies at EU-level. Hence, multilevel governing processes between agencies at both levels will likely emerge, leading to more uniform application and practice of EU regulations. Second, the more administrative capacities are established at the EU level, the stronger the pull effect of EU-level administrative institutions on corresponding national-level institutions. One consequence is that government bureaucrats may carry double-tasked roles in pursuing public governance. Double-tasked government officials personalise multilevel administrative systems by working within national ministries and agencies while partaking in EU administrative networks and interacting with the EU-level executive branch of government.[11]

---

[7] E-I Dumitrescu and others, "Machine Learning or Econometrics for Credit Scoring: Let's Get the Best of Both Worlds" (15 January 2021) <https://papers.ssrn.com/abstract=3553781> (accessed 26 March 2024).

[8] K Langenbucher, 'Responsible A.I.-Based Credit Scoring – A Legal Framework' (2020) 31 European Business Law Review <https://kluwerlawonline.com/api/Product/CitationPDFURL?file=Journals\EULR\EULR2020022.pdf> (accessed 26 March 2024).

[9] G Dimitropoulos and P Hacker, "Learning and the Law: Improving Behavioral Regulation from an International and Comparative Perspective" (2016) 25 Journal of Law and Policy 473.

### III. The AIA implementation and enforcement: the tasks of the Commission

Implementing the AIA and its enforcement involves several non-legislative acts primarily under the Commission's authority according to the EU's rules for implementing powers, the so-called committee procedure[12] (Recital 86 AIA).[13] A notable initial step in this process was the establishment of the AI Office, formalised by the Commission's Decision on 24 January 2024. The remaining steps that must be taken by the Commission to implement and enforce the AIA are summarised in Table 1 and described in more detail subsequently.

#### III.1. Procedures

The European Commission is required to engage with Member State experts and representatives when adopting implementing and delegated acts to ensure the consistent application and detailed implementation of EU laws. Implementing acts aim to apply EU laws consistently across Member States without altering the law (Article 291 TFEU). In contrast, delegated acts are designed to supplement or modify non-essential elements of legislative acts, adding details needed for their implementation (Article 290 TFEU). Implementing acts, governed by the comitology procedure, involve collaboration with a committee of Member State representatives. Under the AIA, this engagement involves only the European AI Board. Delegated acts require consultation with Member State experts but do not involve a formal committee.[14] Delegated acts are subject to scrutiny by the European Parliament and the Council, which have two months to raise objections; otherwise, the Act is adopted. The Commission's powers under the AIA, including adopting delegated acts, are granted for five years and can be silently renewed unless opposed by the European Parliament (EP) and Council (Article 73 AIA). The Commission must keep the EP and Council informed about delegated acts and report on its activities within nine months, allowing for oversight and potential revocation of its powers. Additionally, the

---

[10] An agency is an administrative body that is formally separated from a ministerial, or cabinet-level, department and that carries out public tasks at a national level permanently, is staffed by public servants, is financed mainly by the state budget, and is subject to public legal procedures. Agencies are supposed to enjoy some autonomy from their respective ministerial departments about decision-making. Over time, agencies tend to be moved out of and into ministerial departments, often cyclically (Bach and Jann 2010; Verhoest et al. 2012).

[11] J Trondal, *An Emergent European Executive Order* (Oxford University Press 2010).

[12] Regulation (EU) No 182/2011 dated 16 February 2011.

[13] GJ Brandsma and J Blom-Hansen, *Controlling the EU Executive? The Politics of Delegation in the European Union* (Oxford University Press 2017).

[14] P Craig, *Delegated and Implementing Acts* (1st edn, Oxford University Press 2018) <https://ora.ox.ac.uk/objects/uuid:f8aa03ad-60c1-4415-b5b5-0863b005407d> (accessed 26 February 2024).

**Table 1.** Tasks and responsibilities of the Commission in implementing and enforcing the AIA

| Key aspects | Tasks and responsibilities of the Commission |
| --- | --- |
| a) Procedures | - Establish and work with the AI Office and AI Board to develop implementing and delegated acts<br>- Conduct the comitology procedure with Member States for adopting and implementing acts<br>- Manage delegated act adoption, consulting experts, and undergoing scrutiny by EP and Council |
| b) Guidelines | - Issue guidelines on applying the definition of an AI system and classification rules for high-risk systems<br>- Create risk assessment methods for identifying and mitigating risks<br>- Define rules for "significant modifications" that alter the risk level of a high-risk system |
| c) Classification | - Update Annex III to add or remove high-risk AI system use cases through delegated acts<br>- Classify GPAI as exhibiting "systemic risk" based on criteria like FLOPs and high-impact capabilities<br>- Adjust regulatory parameters (thresholds, benchmarks) for GPAI classification through delegated acts |
| d) Prohibited Systems | - Develop guidelines on AI practices that are prohibited under Article 5 (AIA)<br>- Set standards and best practices to counter manipulative techniques and hazards<br>- Define criteria for exceptions to prohibitions, e.g., for law enforcement use of real-time remote biometric identification |
| e) Harmonised standards and high-risk obligations | - Define harmonised standards and obligations for high-risk system providers, including in-door risk management system (Article 9 AIA)<br>- Standardise technical documentation requirements and - update Annex IV via delegated acts as necessary<br>- Approve codes of practice (Article 56(6) AIA) |
| f) Information and Transparency | - Set information obligations for providers of high-risk systems throughout the AI value chain<br>- Issue guidance to ensure compliance with transparency requirements, especially for GPAI |
| g) Enforcement | - Clarify the interplay between the AIA and other EU legislative frameworks<br>- Regulate regulatory sandboxes and supervisory functions<br>- Oversee Member State setting of penalties and enforcement measures that are effective, proportionate, and deterrent |

Commission is tasked with publishing guidelines and making binding decisions to implement the AIA effectively. The AI Office will support the adoption of implementing and delegated acts, while the AI Board focuses on implementing acts (see Section III).

### III.2. Guidelines operationalising the risk-based approach

The Commission develops guidelines and updates them to assist in implementing the AIA's risk-based approach, focusing on classifying high-risk AI systems (Article 6(5) AIA). Additionally, the Commission uses delegated acts to update Annex III, either adding new high-risk AI use cases or removing ones that no longer pose significant risks, based on criteria such as likelihood of use, autonomy, human oversight, and outcome reversibility, ensuring that these adjustments do not compromise the EU's health, safety, and rights standards (Article 7 AIA).

Considering the risk-based classification of AI systems, which potentially offers a robust regulatory approach by building in regulatory flexibility and applies to general-purpose AI (GPAI, also known as foundation models) albeit under a distinct terminology – namely, the 'high impact capabilities' (Article 51(1) AIA) – these guidelines should also detail methodologies for risk assessments.[15]

Significantly, within this framework, the Commission must define the rules about "significant modifications" that alter the risk level of a (high-risk) system once it has been introduced to the market or put into use (Articles 25(1) and 3(23)). These alterations, not anticipated or accounted for in the initial conformity assessment conducted by the provider, may require the system to be reclassified (Article 96(1), AIA). This involves specifying what amounts to a significant change and outlining the procedures for performing a new conformity assessment (Article 43(4) AIA). Importantly, delineating a significant modification must refer to the purpose of these sections of the AIA, specifically, hedging some specific risks of AI systems in the light of fundamental rights. Hence, only a noticeable, clear, and relevant change to the system's risks – such as discrimination, opacity, unforeseeability, privacy, or the environment – can be a significant modification, in our view. It follows that a standard fine-tuning exercise of foundation models should not lead to a substantial modification unless the process involves, explicitly, particularly biased data sets, the removal of safety layers, or other actions clearly entailing novel or increased risks.[16]

A complementary yet potentially synergistic approach is to adopt pre-determined change management plans akin to those in medicine.[17] These plans are comprehensive documents outlining anticipated modifications to an AI system – covering aspects like model performance adjustments, data inputs, and shifts in intended use – and the methods for assessing such changes. They might establish a proactive accountability methodology[18] for identifying risks and devising mitigation strategies, ensuring modifications align with fundamental rights and AIA goals. Regulators would evaluate these plans during the AI technology's premarket assessment, allowing post-market changes to be efficiently implemented according to the pre-approved plan. Such change management plans do not amount to a substantial modification in the sense of the AIA as they are not unforeseen or unplanned (Article 3(23), AIA). Hence, they afford the distinct advantage of obviating the need for reclassification and a new conformity assessment. However, they cannot capture dynamic and spontaneous changes by developers or deployers.

### III.3. Classification of GPAI

The Commission has notable authority under the AIA to classify GPAI as exhibiting 'systemic risk' (Article 51 AIA).[19] This distinction, establishing the famous two-tiered approach to the regulation of GPAI,[20] is crucial: only systemically risky GPAIs are subject to the more far-reaching AI safety obligations concerning evaluation and red teaming, comprehensive risk assessment and mitigation, incident reporting, and cybersecurity (Article 55 AIA). The classification authority is delineated in Article 51 AIA, which outlines the criteria according to which a GPAI is considered to exhibit systemic risk. The decision to classify a GPAI as systemically risky can be initiated by the Commission itself or in response to a qualified alert from the Scientific Panel confirming the presence of such high-impact capabilities.

---

[15] C Novelli and others, "AI Risk Assessment: A Scenario-Based, Proportional Methodology for the AI Act" (2024) 3 Digital Society 13; C Novelli and others, "Taking AI Risks Seriously: A New Assessment Model for the AI Act" (2023) AI & SOCIETY <https://doi.org/10.1007/s00146-023-01723-z> (accessed 21 July 2023).

[16] C Novelli and others, "Generative AI in EU Law: Liability, Privacy, Intellectual Property, and Cybersecurity" (14 January 2024) <https://papers.ssrn.com/abstract=4694565> (accessed 28 March 2024).

[17] KN Vokinger and U Gasser, "Regulating AI in Medicine in the United States and Europe" (2021) 3 Nature Machine Intelligence 738; J Morley and others, "Governing Data and Artificial Intelligence for Health Care: Developing an International Understanding" (2022) 6 JMIR Formative Research e31623.

[18] C Novelli, M Taddeo and L Floridi, "Accountability in Artificial Intelligence: What It Is and How It Works" (2023) AI & SOCIETY <https://doi.org/10.1007/s00146-023-01635-y> (accessed 31 July 2023).

[19] What counts as a systemic risk in this field is stated at art. 51, point 1 AIA.

[20] P Hacker, A Engel and M Mauer, "Regulating ChatGPT and Other Large Generative AI Models" (arXiv, 10 February 2023) <http://arxiv.org/abs/2302.02337> (accessed 14 February 2023).

The Commission may dynamically adjust regulatory parameters – such as thresholds, benchmarks, and indicators – through delegated acts. The adaptive mechanism is essential for a robust governance model as it ensures that regulations remain relevant amidst the fast pace of technological advancements, including algorithm improvements and hardware efficiency. The capacity to refine these regulatory measures is particularly vital as the trend in AI development moves towards creating more powerful yet "smaller" models that require fewer floating-point operations (FLOPs).[21]

Against this background, Article 52 outlines a process allowing GPAI providers to contest the Commission's classification decisions. This provision is pivotal, potentially becoming a primary area of contention within the AI Act, akin to the legal disputes observed under the DSA, where entities like Zalando and Amazon have disputed their categorisation as Very Large Online Platforms.[22] Particularly, GPAI providers whose models are trained with fewer than $10^{25}$ FLOPs yet are deemed systemically risky are expected to use this mechanism actively, possibly leading to legal challenges that could reach the Court of Justice of the European Union (CJEU). The legal recourse presented here is a double-edged sword. On the one hand, it constitutes an essential component of the AIA, offering a counterbalance to the Commission's regulatory powers and ensuring a venue for addressing potential methodological errors or disputes over classifications. On the other hand, it provides a venue for providers with deep pockets to delay the application of the more stringent rules for systemically relevant GPAI. Simultaneously, this reinforces the importance of the presumptive $10^{25}$ FLOP threshold – which is outdating rapidly due to the growing capabilities of smaller foundation models.

### III.4. Prohibited systems

The Commission is tasked with developing guidelines to address prohibited AI practices (Article 5, AIA), including setting technical standards and best practices for AI system design to prevent manipulative techniques. It must also define criteria for exceptions where AI can be used to address significant threats or terrorist activities, with specific allowances for law enforcement, such as the use of real-time remote biometric identification in public spaces. These guidelines will also outline necessary procedural safeguards to ensure such exceptions do not infringe on fundamental rights. They will be crucial to balance law enforcement needs with individual privacy and freedom protections.[23]

### III.5. Harmonised standards and high-risk obligations

The Commission must also set harmonised standards and define obligations for providers of high-risk AI systems under the AIA, requiring a comprehensive "in-door" risk management process that is continuous and iterative throughout the system's lifecycle.

---

[21] S Ma and others, "The Era of 1-Bit LLMs: All Large Language Models Are in 1.58 Bits" (arXiv, 27 February 2024) <http://arxiv.org/abs/2402.17764> (accessed 28 March 2024).

[22] FY Chee and FY Chee, "Amazon Makes First Big Tech Challenge to EU Online Content Rules" *Reuters* (11 July 2023) <https://www.reuters.com/technology/amazon-challenges-eu-online-content-rules-says-unfairly-singled-out-2023-07-11/> (accessed 28 March 2024).

[23] Notably, a recent CJEU Decision (Case-588/21) mandates the public disclosure of harmonised technical standards to reinforce principles of the rule of law and free access to the law.

This includes detailing timelines, design choices, data processing methods, and strategies to mitigate biases, alongside standardising technical documentation as per Annex IV, with updates via delegated acts to adapt to technological advances and ensure compliance with regulatory standards.

### III.6. Information and transparency

The Commission is also responsible for setting forth information obligations along the AI value chain that reflect the current technological standards for providers of high-risk systems (Article 28 AIA) and offering guidance to ensure compliance with transparency requirements, which holds particular significance for GPAI (Article 53 AIA). To achieve this, the Commission might, for example, issue directives on properly revealing the use of GPAI across different settings, considering the medium and essence of the content implicated.

### III.7. Overlap with other regulations and enforcement timeline

Finally, the Commission must elucidate the interplay between the AIA and other EU legislative frameworks to guarantee internal systematicity and consistent enforcement across the board. Specifically, the Commission may need to offer illustrative examples of potential overlaps or conflicts and promoting the formation of joint oversight entities or working groups. Such initiatives would facilitate the exchange of information, standardise enforcement approaches, and develop unified interpretative guidelines, ensuring a harmonised regulatory landscape across the European Union.

The AIA's enforcement is structured in stages, with transition periods for compliance varying by the risk level of AI systems and linked to the Act's official entry into force. Specific grace periods are set for different categories of AI systems, ranging from 6 to 36 months. However, for existing GPAI systems already on the market, a grace period of 24 months is granted before they must comply fully (Article 83(3) AIA). Even more importantly, high-risk systems already on the market 24 months after the entry into force are entirely exempt from the AIA until significant changes are made in their designs (Article 83(2) AIA). Conceptually, this meaningful change can be equated with the substantial modification discussed above. Arguably, however, this blank exemption is in deep tension with a principle of product safety law: it applies to all models on the market, irrespective of when they entered the market. Moreover, the grace period for GPAI and the exemption for existing high-risk systems favor incumbents vis-à-vis newcomers, which is questionable from a competition perspective.

## IV. Supranational authorities: the AI Office, the AI Board, and the other bodies

The AIA mandates a comprehensive governance framework, as highlighted in Recital 5 of the Commission's Decision that establishes the AI Office. Under this framework, the AI Office is responsible for overseeing AI advancements, liaising with the scientific community, and playing a pivotal role in investigations, testing, and enforcement, all while maintaining a global perspective.

The governance structure proposed by the AIA involves establishing national and supranational bodies. Two key institutions are formed at the supranational level: the AI Office and the European AI Board. While distinct in structure and task, these entities are somehow complementary. The AI Office is anticipated to focus on regulatory oversight and enforcement, especially concerning GPAI models. The European AI Board is expected to ensure coordination among Member States, enhancing the AIA's implementation through advice, consultation, and awareness initiatives. Besides these two, the AIA also introduces other significant, though partially autonomous, supranational bodies, namely the Scientific Panel and the Advisory Forum.

Table 2 below outlines the structure, composition, missions, and tasks of the institutional bodies implementing and enforcing the AIA. It provides a foundation for the more detailed discussion to follow in subsequent sections.

### IV.1. The AI Office

The first step in implementing the AIA was establishing a centralised AI Office, in January 2024.[24] Its primary mission is to lay down harmonised rules to implement and enforce the AIA consistently across the EU. The formation of the AI Office is geared towards unifying Europe's AI expertise by leveraging insights from the scientific domain. In implementing the AI Act, much will depend on "getting the AI Office right."

The Office's broad mandate involves collaboration with scientific experts, national authorities, industry representatives, and significant institutions like the European High-Performance Computing Joint Undertaking and international organisations. An important aspect of the AI Office's role is overseeing General-Purpose AI (GPAI) technologies, exemplified by ChatGPT and Gemini (e.g., Articles 52 to 56 AIA).

#### a. Institutional identity, composition, and operational autonomy

Regarding its institutional identity, the AI Office resembles EU interinstitutional services, marked by its focused scope – currently dedicated solely to implementing the AIA – and its role in providing cross-support to various institutions such as the EP, Council, and the Central Bank. Like interinstitutional services, it extends support to agencies and bodies like the European Data Protection Board and the European Investment Bank. It is explicitly stipulated in Articles 5 and 6 of the Commission's Decision that the AI Office is entrusted with supporting the European Artificial Intelligence Board and collaborating with the Centre for Algorithmic Transparency. This places the AI Office alongside other interinstitutional services, such as the Computer Emergency Response Team (CERT-EU), illustrating its distinctive function within the EU framework.

However, unlike interinstitutional services, the AI Office is integrated within the administrative framework of a single entity, specifically the DG for Communication Networks, Content, and Technology (DG-CNECT) of the Commission. The AI Office thus represents primarily a centralised institutional design (see Option 1 above). DG-CNECT operates similarly to a national ministry, overseeing the implementation of policies and programs related to the digital single market. Within DG-CNECT, there are multiple units (called Connects), each specialising in various facets of digital policy, technology, and administration. These units often have overlapping competencies, and the AI Office engages in cross-cutting issues relevant to several of them, with Connect A ('Artificial Intelligence and Digital Industry') being particularly central.

Integrating the AI Office within the DG implies that it operates under the regulations and procedural framework of the Commission. However, the AI Office's precise organisational structure, specific method of ensuring expertise, and operational autonomy remain ambiguous. No provisions, either in the AIA or in the Commission's Decision, have been established regarding the composition of the AI Office, its collaborative dynamics with the various Connects within the DG, or the extent of its operational autonomy. The AIA emphasises the necessity for national competent authorities to possess "adequate

---

[24] It has been established through a Commission's Decision (Brussels, 24.1.2024, C (2024) 390 final).

*European Journal of Risk Regulation* 11Table 2. Structures, compositions, missions, and tasks of the institutional bodies involved in the AIA implementation and enforcement

| Institutional Body | Structure and Composition | Mission and Tasks |
| --- | --- | --- |
| AI Office (Art. 64 AIA and Commission's Decision) | Centralised within the DG-CNECT of the Commission | - Harmonise AIA implementation and enforcement across the EU<br>- Support implementing and delegated acts<br>- Standardisation and best practices<br>- Assist in the establishment and operation of regulatory sandboxes<br>- Assess and monitor GPAIs and aid investigations into rule violations<br>- Provide administrative support to other bodies (Board, Advisory Forum, Scientific Panel)<br>- Consult and cooperate with stakeholders<br>- Cooperate with other relevant DG and services of the Commission<br>- International cooperation |
| AI Board (Art. 65 AIA) | Representatives from each Member State, with the AI Office and the European Data Protection Supervisor participating as observers | - Facilitate consistent and effective application of the AIA<br>- Coordinate national competent authorities<br>- Harmonise administrative practices.<br>- Issue recommendations and opinions (upon requests of the Commission)<br>- Support the establishment and operation of regulatory sandboxes<br>- Gather feedback on GPAI-related alerts |
| Advisory Forum (Art. 67 AIA) | Stakeholders appointed by the Commission | - Provide technical expertise<br>- Prepare opinions and recommendations (upon request of the Board and the Commission)<br>- Establish sub-groups for examining specific questions<br>- Prepare an annual report on activities |
| Scientific Panel (Art. 68 AIA) | Independent experts selected by the Commission | - Support enforcement of AI regulation, especially for GPAI<br>- Provide advice on the classification of AI models with systemic risk<br>- Alert AI Office of systemic risks<br>- Develop evaluation tools and methodologies for GPAIs<br>- Support market surveillance authorities and cross-border activities |
| Notifying Authorities (Artt. 28-29 AIA) | Designated or established by Member States | - Process applications for notification from conformity assessment bodies (CABs)<br>- Monitor CABs<br>- Cooperate with authorities from other Member States<br>- Ensure no conflict of interest with conformity assessment bodies<br>- Conflict of interest prevention and assessment impartiality |

(*Continued*)https://doi.org/10.1017/err.2024.57 Published online by Cambridge University Press

**Table 2.** (*Continued*)

| Institutional Body | Structure and Composition | Mission and Tasks |
| --- | --- | --- |
| Notified Bodies (Artt. 29-38 AIA) | A third-party conformity assessment body (with legal personality) notified under the AIA | - Verify the conformity of high-risk AI systems<br>- Issue certifications<br>- Manage and document subcontracting arrangements<br>- Periodic assessment activities (audits)<br>- Participate in coordination activities and European standardisation |
| Market Surveillance Authorities (Artt. 70–72 AIA) | Entities designated or established by Member States as single points of contact | - Non-compliance investigation and correction for high-risk AI systems (e.g., risk measures)<br>- Real-world testing oversight and serious incident report management<br>- Guide and advice on the implementation of the regulation, particularly to SMEs and start-ups<br>- Consumer protection and fair competition support |

technical, financial, and human resources, and infrastructure to fulfill their tasks effectively, with a sufficient number of personnel permanently available" (Article 70(3) AIA), covering expertise areas from data computing to fundamental rights. However, there are no equivalent stipulations for the AI Office. This absence is likely justified by the expectation that the AI Office will, at least partially, use the existing infrastructure and human resources of the DG-CNECT. Nonetheless, expert hiring and substantial funding will be crucial for the success of the Office, an issue that represents a significant challenge for the public sector as it competes with some of the best-funded private companies on the planet.

Regarding its operational autonomy, the AI Office is subject to two primary constraints, aside from its absence of legal personality, which sets it apart from EU agencies. First, its incorporation into the administrative structure of DG-CNECT means that DG-CNECT's management plan will guide the AI Office's strategic priorities and the distribution of resources, directly influencing the scope and direction of its initiatives.

Second, the operational autonomy of the AI Office is further restricted by the defined competencies of other entities, including EU bodies, offices, agencies, and national authorities. While it seems appropriate for the AI Office to perform its duties in issuing guidance without duplicating the efforts of relevant Union bodies, offices, and agencies under sector-specific legislation (as per Recital 7 of the Commission's Decision), the mechanisms for coordination remain ambiguous. This ambiguity includes how conflicts or overlaps in competencies – e.g., with the European Data Protection Board (EDPB) concerning data quality and management obligations for providers of high-risk systems as stipulated by the AIA – will be managed. Consider the development of a healthcare AI system handling sensitive personal data. Here, the AI Office may emphasise the system's innovative contributions to healthcare. In contrast, the EDPB might insist on strict adherence to GDPR data protection principles, potentially causing tensions in the system's deployment and usage.

Resolving such discrepancies could involve defining the AI Office's organisational structure and operational scope. Clarification could include specifying whether collaborative mechanisms exist, such as joint working groups between the AI Office and other EU entities or establishing interagency agreements. Such agreements could mirror the Memorandum of Understanding between the European Data Protection Supervisor (EDPS) and the European Union Agency for Cybersecurity (ENISA).

*b. Mission(s) and task*

While the DG-CNECT pursues a wide range of goals from internet governance to green development, the AI Office's primary mission, according to the Commission Decision, is to ensure the harmonised implementation and enforcement of the AIA (Article 2, point 1 of the Decision). However, the Decision outlines auxiliary missions: enhancing a strategic and effective EU approach to global AI "initiatives", promoting actions that maximise AI's societal and economic benefits, supporting the swift development and deployment of trustworthy AI systems that boost societal welfare and EU competitiveness, and keeping track of AI market and technology advancements (Article 2, point 2).

The language used in the provisions concerning the AI Office tasks, notably in Article 2a, is broad and open-ended, referring to contributions to "initiatives on AI" without specifying details. Ambiguity may have been intentional, inviting further interpretation. One interpretation is that the AI Office's role could go beyond the scope of the AIA to include support for implementing additional AI normative frameworks, such as the revised Product Liability Directive or the Artificial Intelligence Liability Directive (AILD). Expanding the AI Office's remit in this way could be advantageous, as limiting its focus to a single regulation might lead to squandering valuable expertise developed through the AIA's implementation. Moreover, broadening the AI Office's mandate to ensure harmonising the AIA's rules with other AI regulations could prevent conflicts and inconsistencies, thereby aiding Member States and their respective authorities in adopting a comprehensive AI legislative framework. Such an expansion suggests the Office's potential as an emerging EU "digital agency".

However, designating the AI Office as the competent authority for multiple regulatory frameworks – with varying tasks depending on the specific framework – points to a potential need for restructuring. The AI Office might require future transformation into a more autonomous body. Without necessarily acquiring legal personality, the AI Office might evolve into an inter-institutional service like the CERT-EU, which could necessitate detaching it from the Commission's administrative framework.

The main issue with transforming the AI Office into an inter-institutional service lies in the inherent design of such services. They are primarily established to offer widespread support across EU institutions, focusing on internal functionalities such as recruitment (via the European Personnel Selection Office, EPSO), staff training, promoting inter-institutional collaboration, and facilitating the efficient execution of legislative and policy frameworks. In contrast, the AI Office's mandate involves spearheading the implementation of the AIA, entailing the issuance of guidelines, regulation enforcement, and compliance oversight. Given such entities' predominantly supportive and non-regulatory nature, transitioning into an inter-institutional service might dilute its capability to perform these critical functions.

Against this background, a crucial aspect concerning the AI Office is the ambiguity in the current normative framework regarding the breadth of its mission scope. This ambiguity sets it apart from the European Food Safety Authority (EFSA) or the European Medicines Agency (EMA) (in terms of isolation problems) but also opens the door to a potentially beneficial interpretation, allowing the AI Office to oversee multiple entities. It is critical to emphasise that adopting this more comprehensive interpretative approach would require appropriate changes in the institutional design to accommodate the Office's extended functions.

Concerning its specific tasks, the AI Office plays a pivotal role in applying and enforcing regulations concerning GPAI, focusing on standardisation efforts to harmonise tools, methodologies, and criteria for evaluating systemic risks associated with GPAI across supranational and national levels. It also monitors GPAIs continuously for adherence to standards and potential new risks, supports investigations into GPAI violations, and assists in developing delegated acts and regulatory sandboxes for all AI systems under the AIA.

The responsibilities assigned to the AI Office are broadly defined, with the expectation that their precise implementation will evolve based on practical experience. The AI Office requires significant expertise and financial resources to offer support and technical advice across diverse tasks and AI systems. Achieving this will demand a robust administrative framework that effectively manages internal subgroup coordination and external engagements with supranational and national entities.

An important aspect to consider within the operational scope of the Office is the nature of its decisions. The AI Office does not issue binding decisions on its own. Instead, it provides support and advice to the Commission. Nonetheless, it plays a role in formulating the Commission's decisions, including implementing and delegated acts, which, while non-legislative, are binding across all Member States. These decisions by the Commission can be challenged based on various grounds, such as exceeding its authority or misusing its powers, and through specific processes before and after they are formally adopted. For example, before adoption, implementing and delegated acts can be contested through feedback mechanisms provided by committees (as part of the comitology) or by EU institutions; once adopted, these acts are subject to judicial review by the Court of Justice of the EU, which assesses their compliance with the foundational legislation.[25]

However, the effectiveness of mechanisms for appealing decisions may be compromised by the opaque nature of the AI Office's support to the Commission, its interactions within DG-CNECT, and its relationships with external bodies, such as national authorities. Such opacity can obscure the reasons behind certain implementing or delegated acts, which is particularly concerning given the AI Office's engagement with external experts and stakeholders.[26] Accordingly, the documentation and disclosure of the AI Office's contributions, as evidenced through summary records in the comitology register and the explanatory memoranda accompanying the Commission's delegated acts, become crucial.

### IV.2. The AI Board, the Advisory Forum, and the Scientific Panel

The European Artificial Intelligence Board (hereafter "the Board") is distinct from the AI Office. Yet, it undertakes tasks that are parallel and intersect with those of the AI Office, particularly in supervising and directing the execution of the AIA. Currently, the governance and operational structure of the Board is primarily detailed in Articles 65 and 66 of the AIA. In addition to the Board, the AIA also establishes other bodies that, while independent in their formation, support the Board: the Advisory Forum and the Scientific Panel. The result is a complex network of bodies, making their coordination challenging.

#### a. Structures, roles, and composition of the three bodies

The Board consists of a representative from each Member State, appointed for three years, with one of them as the chair. The AI Office and the European Data Protection Supervisor participate as observers without voting powers. Unlike the generic recruitment criteria for the AI Office, the AIA explicitly requires that Member States appoint representatives to the Board who have the requisite expertise and authority in their respective countries to contribute to the Board's missions effectively. These representatives are also empowered to gather essential data and information to ensure uniformity and coordination among national competent authorities (Article 65(4)(c) AIA). Coordination is supported by two permanent sub-groups, which serve as platforms for collaboration and information sharing between market surveillance and notifying authorities. Additionally, the Board has the authority to form temporary sub-groups to delve into other specific topics.[27]

---

[25] R Dehousse, "Comitology: Who Watches the Watchmen?" (2003) 10 Journal of European Public Policy 798; Brandsma and Blom-Hansen (n 14).

[26] This also emerges from the published call for interests: https://digital-strategy.ec.europa.eu/en/policies/ai-office.

The Board may invite other authorities or experts on a case-by-case basis. However, it will be supported by an Advisory Forum (hereafter "the Forum"), which provides technical expertise also to the Commission (Article 67 AIA). The Commission will ensure the Forum includes diverse stakeholders, such as industry representatives, start-ups, small and medium-sized enterprises (SMEs), civil society groups, and academic institutions, to offer comprehensive stakeholder feedback to the Commission and the Board.

Finally, the AIA mandates the creation of a Scientific Panel of independent experts (hereafter "the Panel") through a Commission implementing act, aimed at bolstering the AIA's enforcement activities (Article 68 AIA). In consultation with the Board, the Commission will determine the Panel's membership, selecting experts based on their specialised knowledge and independence from AI system providers. The Panel is designed to be a resource for Member States and assist them in enforcing the AIA. It should be noted that Member States may need to pay fees for the expert advice and support provided by the Panel (Article 69 AIA).

Within the EU's regulatory framework for AI, the AI Board operates as an advisory body, the Advisory Forum acts as a consultative body offering industry insights to both the Board and the Commission, and the Scientific Panel primarily provides expert scientific support to the AI Office and the Member States in need of its specialised knowledge. The composition of these entities varies: the members of the AI Board are appointed by Member States, and the Commission and the Board choose the Advisory Forum's members. In contrast, the Scientific Panel's members are appointed solely by the Commission.

Despite the distinct tasks assigned to them, which will be discussed later, the necessity of having three separate entities with pretty similar compositions raises questions. The AI Board's establishment is understandable for ensuring representation and coordination among Member States and maintaining some independence from EU institutions, without requiring members to possess scientific expertise. However, the rationale behind keeping the Advisory Forum *and* the Scientific Panel is not apparent. The Advisory Forum is intended to draw upon the diverse perspectives of civil society and industry sectors, essentially acting as an institutionalised form of lobbying to represent their varying commercial interests. However, it must also ensure a balance with non-commercial interests. In contrast, the Scientific Panel consists of independent and (hopefully) unbiased academic experts with specific tasks related to GPAIs.

### b. Mission(s) and tasks: Ockham's razor

The three bodies are set to perform slightly different tasks. The Board undertakes numerous tasks (Article 66 AIA), such as providing guidance and support to the Commission and Member States to coordinate national authorities. It offers recommendations for delegated and implementing acts and aims to standardise administrative practices across Member States, for instance, through addressing exemptions from conformity assessment procedures and by supporting the operation of regulatory sandboxes (Article 66(d) AIA). Additionally, the Board advises on creating codes of conduct and applying harmonised standards, supports the AI Office in helping national authorities establish and enhance regulatory sandboxes, and gathers feedback from Member States on alerts related to GPAIs.

---

[27] The AIA has outlined initial functions and roles for the Board, yet additional details and responsibilities are expected to be further delineated in subsequent legislation.

Note that, other than the European Data Protection Board under Article 65 GDPR, the AI Board does not have the authority to revise national supervisory agency decisions or resolve disputes between national authorities with binding force. Under the GDPR, this has emerged as a critical mechanism, particularly in dealing with the contentious ruling of the Irish Data Protection Commission concerning big technology companies headquartered in Ireland.[28] This lack of a corresponding authority for the AI Board may prove a distinct disadvantage, hindering the uniform application of the law, if some Member States interpret the AIA in highly idiosyncratic fashions (as the Irish Data Protection Commission did with the GDPR). In this context, one may particularly think of the supervision of the limitations enshrined in Article 5 AIA on surveillance tools using remote biometric identification. European oversight may be required, especially in countries with significant democratic backsliding, to avoid the abuse of AI for stifling legitimate protest and establishing an illiberal surveillance regime.

The Board and Office collaboration will be characterised by mutual support. However, while there are areas where the functions of the AI Board and the AI Office might seem to overlap, especially from the viewpoint of Member States, merging these two entities is not viable. The need for political representation and their distinct roles – where the Board provides advisory insights and the Office executes the Commission's binding decisions – prevents such a merger.

Similarly, the roles of the Advisory Forum and the Scientific Panel might not be distinctly demarcated. The Advisory Forum has a broad mandate to provide advice and expertise to the Board and the Commission, supporting various tasks under the AIA. The Scientific Panel advises and supports the AI Office, specifically on implementing and enforcing the AIA, focusing on GPAIs. Its tasks include developing evaluation tools, benchmarks, and methodologies for GPAIs, advising on classifying GPAIs with systemic risks, and assisting Member States in their enforcement activities as requested (Article 68 AIA).

However, the distinction between the Advisory Forum and the Scientific Panel seems less precise than the separation between the Board and the Office. Questions arise regarding the exclusivity of the Forum's support to the Board and the Commission and whether the Panel's specialised GPAI expertise could benefit these entities. If the overarching aim of the AIA's governance structure is to secure impartial and external feedback for the comprehensive implementation and enforcement, then such support should be accessible to all EU institutional bodies involved – namely, the Commission and its Office – as well as to Member States, whether through the Board or their respective national authorities. While it might be argued that the AI Office's participation in Board meetings is an indirect channel for the Panel's expertise to influence broader discussions, this arrangement is not entirely satisfactory. The indirect nature of this influence means that the Panel's specialised opinions could become less impactful, especially since the AI Office's contributions to the Board's meetings lack formal voting power, which could further dilute the Panel's input. In any case, the fact that the Panel's insights are indirectly presented to the Board is an *a fortiori* argument against the continued separation of the Advisory Forum and the Scientific Panel. If the separation was meant to distinguish the types of support provided by each entity, indirect participation blurs these lines. This topic will be expanded upon in section VI.

---

[28] R Boardman, "Key Takeaways from the Irish DPC and EDPB Decisions on Facebook Data Transfers, DPC Decision, EU User Data" *Bird&Bird* (23 May 2023) <https://www.twobirds.com/en/insights/2023/global/key-takeaways-from-the-irish-dpc-and-edpb-decisions-on-facebook-data-transfers> (accessed 29 March 2024).

## V. National authorities: Notifying authorities, notified bodies, and market surveillance authorities

Supranational authorities have an essential role, but the effective implementation and enforcement of this Regulation frequently require a local presence, placing the responsibility primarily on Member States. Each is expected to set up at least one notifying authority responsible for compliance and certification processes and one market surveillance authority to verify that products meet EU harmonisation legislation standards for safety, health, and environmental protection as outlined in Regulation (EU) 2019/1020.[29] Both authorities are also encouraged to guide compliance to SMEs and start-ups, considering any relevant recommendations from the Board and the Commission (Chapter VII, Section 2 AIA).

The AIA mandates that national authorities must have permanently available staff with expertise in AI, data protection, cybersecurity, fundamental rights, health and safety, and relevant standards and laws.[30] Member States must assess and report this adequacy to the Commission every two years (Article 70 AIA). Such a requirement illustrates how EU policy regulations can include an organiaational component that interacts with the historical prerogatives of national governments to structure the state apparatus at its own will ("administrative sovereignty").

In this context, Member States have the flexibility to design their governance structures for AI regulation: they can either establish new regulatory bodies dedicated to AI or integrate these oversight responsibilities into existing entities, like national Data Protection Authorities, within their legal frameworks. The autonomy of Member States in this context enables them to delegate tasks to the most suitable public organisations, as discussed above (Part II.b)).

### V.1. Notifying authority and notified bodies

Notifying authorities are national entities established by each Member State to evaluate, designate, and recognise conformity assessment bodies and oversee their activities (Article 28 AIA).

Entities seeking to perform conformity assessments under the AIA must apply to the notifying authority in their Member State or a third country, providing a detailed description of their assessment activities, used modules, AI systems competencies, and an accreditation certificate from a national body. Once an applicant is verified to meet all criteria, the notifying authority endorses it as a notified body, officially recognised to evaluate AI system conformity before market release. Notifying authorities oversee these bodies impartially, are prohibited from engaging in assessment activities to avoid conflicts of interest, and can restrict, suspend, or withdraw a body's status if it fails to meet obligations.

Notified bodies are responsible for impartially and confidentially assessing high-risk AI systems. They ensure that these meet regulatory standards and possess the necessary expertise, including for outsourced work. They have the right to unrestricted access to relevant datasets and may request additional testing to confirm compliance. Upon a satisfactory assessment, they issue an EU technical documentation assessment certificate, valid for up to five years, depending on the AI system category. Notified bodies must justify their certification decisions, which can be appealed by providers, and are required to inform authorities about their certification decisions and significant operational changes, fostering transparency and accountability in the AI certification process.

---

[29] This Regulation's market surveillance targets products under Union harmonisation listed in Annex I, excluding food, feed, medicines, live plants and animals, and reproduction-related products.

[30] The operational details for notifying authorities and notified bodies are specified in Chapter 4, Title III of the AIA, and guidelines for notifying and market surveillance authorities are in Title VI, Chapter 2 ('Governance').

The regulatory framework governing the structure and operations of notifying authorities and notified bodies looks robust. It is also, to an extent, tried and tested as notified bodies of the European Medicines Agency are also accredited to conduct conformity assessments for medical devices. However, it still falls short in terms of specificity, particularly regarding the mechanisms to ensure their impartiality and prevent conflicts of interest (also indirect ones). Impartiality is crucial for upholding the integrity of the conformity assessment process. The AIA stipulates that notifying authorities must be organised and function to avoid conflicts of interest with conformity assessment bodies (Article 30 AIA). However, it lacks detailed guidance on the implementation of such measures. It does not explicitly designate who is responsible for enforcing these requirements – as instead does with the AI Office for the Scientific Panel (Article 68 AIA) – particularly regarding establishing effective oversight mechanisms like regular audits. While the supervision conducted by accreditation bodies does provide some level of oversight, focusing mainly on the competence and compliance with quality standards (such as ISO) of the notified bodies, these controls may not be comprehensive enough. They tend to concentrate on these bodies' technical competencies and quality management systems rather than addressing the broader issues of ensuring impartiality and avoiding conflicts of interest.

Literature raises two main concerns about notified bodies. First, there is a lack of organisational and operational transparency, a situation worsened by these bodies frequently outsourcing their tasks.[31] Second, there are significant concerns about the neutrality of these notified bodies due to their financial relationships with AI providers. These relationships, which can involve fees or commissions, might compromise their decision-making, casting doubt on their ability to regulate effectively.[32]

Another critical aspect requiring attention is the coordination among notified bodies (Article 38 AIA) to prevent divergent interpretations and applications of EU directives and regulations by Member States. Divergences could lead to inconsistencies in how notified bodies are designated and monitored in different jurisdictions. This situation echoes challenges observed in other sectors, such as healthcare, where drugs and devices not approved in one region may seek approval in another. Ultimately, "conformity shopping" must be prevented. Thus, refining and clarifying the regulations concerning the impartiality and oversight of notified bodies is a critical challenge that the existing regulatory framework needs to address more thoroughly.

### V.2. Market surveillance authority

Under the AIA, Member States are mandated to appoint a specific Market Surveillance Authority (MSA) to serve as a single point of contact (Art. 70 AIA) and to communicate the designated point of contact to the Commission. Following this, the Commission will provide a list of the single points of contact available to the public.

In the EU, market surveillance ensures that products meet health and safety standards, supporting consumer protection and fair competition through inspections, document reviews, and compliance tests. An EU-wide product compliance network facilitates adherence to standards by encouraging collaboration and information sharing among Member States and customs authorities to uphold product safety and integrity.[33]

---

[31] J-P Galland, "The Difficulties of Regulating Markets and Risks in Europe through Notified Bodies" (2013) 4 European Journal of Risk Regulation 365.

[32] A Cefaliello and M Kullmann, "Offering False Security: How the Draft Artificial Intelligence Act Undermines Fundamental Workers Rights" (2022) 13 European Labour Law Journal 542.

Under the AIA, MSAs have enhanced powers to oversee high-risk AI systems, particularly those used in law enforcement. These powers include accessing processed personal data, relevant information, and, if necessary, AI system source codes to verify compliance. MSAs can also bypass standard assessment procedures under exceptional circumstances, such as threats to public security or health, and are involved in real-world testing, managing incident reports, and enforcing risk mitigation measures for compliant AI systems that still pose public threats. The Commission coordinates these efforts through the AI Office. MSAs act as central contact points for administrative and public inquiries facilitated by the EUGO network within the digital e-government platform framework.

Furthermore, Market Surveillance Authorities (MSAs) are tasked with liaising with national public authorities responsible for ensuring compliance with Union laws that protect fundamental rights, including non-discrimination principles, such as national data protection authorities. The AIA grants these authorities the right to demand and access any relevant documentation maintained under the regulation in a format that is accessible and comprehensible (Article 77 AIA). The right to access must be granted when these authorities must effectively carry out their duties within their legal scope. When such documentation is requested, the relevant public authority or body must notify the corresponding MSA in the Member State involved.

A key question regarding MSAs is their institutional design. Article 70 of the AIA allows Member States flexibility in establishing or integrating new authorities into existing ones. In practice, the design choice may be contingent on pre-existing administrative structures of MSAs (see Part 2b). Currently, EU Member States possess various authorities categorised by sector (e.g., Medical Devices, Construction Products, Motor Vehicles) listed on the Commission portal.[34] One approach to minimise the creation of new authorities would be establishing dedicated sub-sections within existing MSAs, granting them exclusive competence over AI used in their specific sectors. However, this strategy might not be sufficient in the long run. Considering the anticipated growth in AI functionalities and (EU) legislation, a dedicated MSA for AI products and services will likely be more appropriate, with the three options discussed above ranging from an entirely new agency to a "competence center" within an existing one (Part 2b).

The AIA also remains unclear on whether users or third parties negatively impacted by AI systems will have the right to complain to MSAs. The GDPR grants individuals the right to file complaints and seek judicial remedies against supervisory authorities. The absence of a similar right for AI-related grievances under the AIA would undermine its safeguard of access to justice.[35]

Another critical aspect regards the need for harmonisation among various MSAs, with a specific concern about disparities in resource allocation by Member States. The AIA underscores the importance of equipping national competent authorities with sufficient technical, financial, and human resources (Article 70(3) AIA). However, discrepancies in the provision of these resources across Member States can lead to uneven enforcement and oversight, with implications for market growth and innovation. A prime example of this is the potential delay in product investigations. A lack of sufficient resources at MSAs can significantly slow the regulatory oversight of emerging AI technologies. For start-ups and companies driven by innovation, the speed at which they can enter the market is crucial.

---

[33] Regulation (EU) 2019/1020 is the cornerstone of the legal framework governing market surveillance authorities in the European Union. It replaces the earlier market surveillance provisions outlined in Regulation (EC) No 765/2008. However, other relevant regulations also play a part, such as Decision 768/2008/EC and Directive 2001/95/EC.

[34] https://single-market-economy.ec.europa.eu/single-market/goods/building-blocks/market-surveillance/organisation_en.

[35] M Fink, "The EU Artificial Intelligence Act and Access to Justice" <https://hdl.handle.net/1887/3180727> (accessed 30 April 2024).

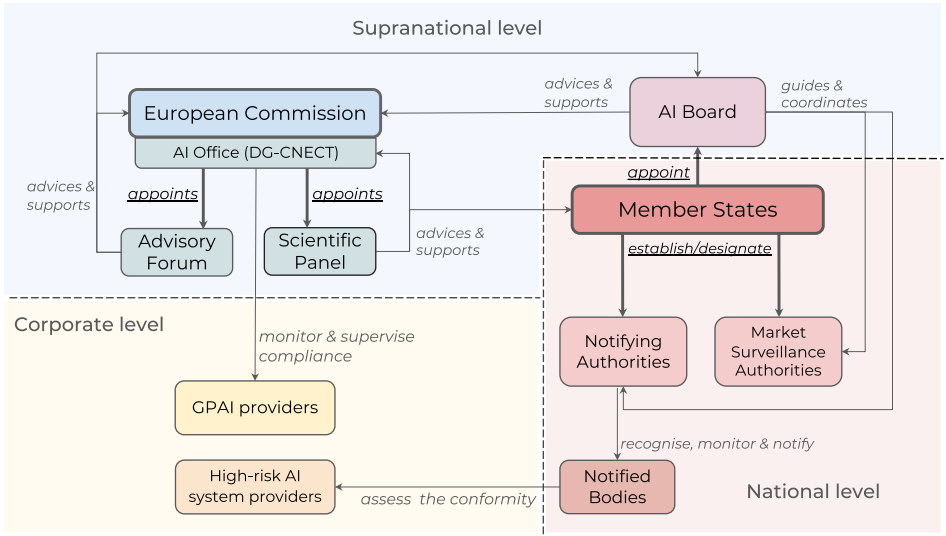

**Figure 1.** Supranational and national bodies involved in the implementation and enforcement of the AIA.

Against this background, it is necessary to strengthen existing coordination mechanisms within the EU, such as the EU Product Compliance Network (EUPCN), established by the Market Surveillance Regulation (2019/1020). Comprising representatives from each EU country, the EUPCN aims to facilitate the identification of shared priorities for market surveillance activities and the cross-sectoral exchange of information on product evaluations. This includes risk assessment, testing methods and outcomes, and other factors pertinent to control activities. It also focuses on the execution of national market surveillance strategies and actions. Such enhanced coordination is vital for mitigating disparities in resource availability and ensuring a more uniform approach to the regulation and oversight of AI technologies across the EU. To conclude this section, in Figure 1, we provide a visual representation to delineate the principal genetic (indicated by a *bold* line) and functional (represented by a *thin* line) connections among the institutional entities engaged in executing and enforcing the AIA.

## VI. Towards simplified and robust governance: recommendations

Building on this analysis, we envision several important updates that should be made to the governance structure of the AI Act.

### VI.1. Clarifying the institutional design of the AI Office

Given the broad spectrum of tasks anticipated for the AI Office – from evaluating GPAIS' capabilities to assisting in creating regulatory sandboxes – more detailed organisational guidance seems needed to identify its institutional design. Additionally, the mandate for the AI Office to "involve independent experts to carry out evaluations on its behalf" (Recital 164 AIA) lacks specificity concerning the criteria for selecting these experts. A more structured approach, similar to the UK's model for health technology assessments where the National Institute for Health and Care Excellence (NICE) sets precise criteria for evidence evaluation and commissions independent reviews from entities like the Cochrane Collaborative, could inform the AI Office's procedures. This model, supported by government funding, provides a structured and standardised method that could inform the AI Office's procedures to ensure its effectiveness in fulfilling its diverse responsibilities.

Another critical consideration is the potential impact of integrating the AI Office within the overarching framework of the Commission, which may obscure its operational transparency. Concern stems from the obligation to adhere to the Commission's general policies on communication and confidentiality. For example, the right to public access to Commission documents, governed by Regulation (EC) No 1049/2001, includes numerous exceptions that could impede the release of documents related to the AI Office. One such exception allows EU institutions to deny access to records if it would compromise the "[...] commercial interests of a natural or legal person, including intellectual property," a broadly defined provision lacking specific, enforceable limits. To mitigate this risk, applying a narrower interpretation of these exceptions to the AI Office would align with recent trends in the case law of the EU Courts,[36] helping to avoid transparency issues similar to those faced by other EU agencies, such as Frontex.[37]

Furthermore, greater clarity regarding the AI Office's operational autonomy is needed. For example, guidelines delineating its decision-making authority, financial independence and engagement capabilities with external parties would be beneficial. As described previously (Recital 14), the call for involving independent experts is a step in the right direction. Still, detailed criteria for expert selection and involvement are required to ensure transparency and efficacy in its evaluation and advisory roles.

An alternative, potentially more effective approach would be establishing the AI Office as a decentralised agency with its legal identity, like the EFSA and the EMA. Decentralisation, designed for pivotal sectors within the single market, would endow the AI Office with enhanced autonomy, including relative freedom from political agendas at the Commission level, a defined mission, executive powers, and the authority to issue binding decisions, albeit with options for appeal and judicial scrutiny. Such an organisational shift would likely boost the AI Office's independence from the Commission and the broader EU institutional matrix, positioning it as a key player in AI governance. However, this change could lead to *agency drift*, where operations by the AI Office conflict with the wishes or strategies of the Commission. However, empirical evidence suggests that the main interlocutors of EU agencies are "parent" Commission DGs. Therefore, despite adopting a decentralised agency format, the AI Office will likely sustain a strong relationship with the Commission.[38] Empirical studies suggest that one effective mitigation strategy against agency drift is establishing organisational units within the Commission that duplicate or overlap those of the agency.[39] This increases the organisational capacities and expertise within the Commission to oversee and control the agency.

The decentralised alternative still carries inherent risks and challenges. A notable concern is the potential for the AI Office to become somewhat isolated from the rest of the EU institutional ecosystem, which could undermine the effectiveness of its supervision and diminish the capacity for cohesive, EU-wide responses and strategies concerning AI regulation. A similar concern is the potential for this isolation to be used strategically to limit workload by, for example, declaring only "pure AI" within the AI Office's remit and any AI tool embedded or interacting with non-AI components outside of scope. Moreover, this structure amplifies concerns related to "agencification," a term critics use to describe the risks of granting regulatory bodies *excessive* but not *complete* autonomy in practice since these are agencies designed as units subordinated to ministry-like institutions in government systems. Such autonomy could lead to their actions diverging from, or complicating, the EU's overarching objectives and governance frameworks. Critics argue that this could result in a democratic legitimacy deficit or undermine the principles that guide how the EU operates and delegates its powers.[40] To mitigate the risks associated with agencification, implementing more robust ex-ante and ex-post evaluation mechanisms for agency performance is advisable (as we suggest in recommendation (d)). These evaluations, maybe conducted periodically by the European Commission, would assess the impact of agency actions and regulations, ensuring alignment with EU objectives and principles. A complimentary, or alternative, mitigation strategy would be to establish *organisational duplication and overlap* within the Commission, as outlined above.

---

[36] A Marcoulli and L Cappelletti, "Recent Trends and Developments in the Case Law of EU Courts on Access to Documents" (2023) 23 ERA Forum 477.

[37] L Salzano and M Gkliati, "Mysteriousness by Design: The Case of Frontex and the Regulation on Public Access to Documents" (30 September 2023) <https://papers.ssrn.com/abstract=4588534> (accessed 23 February 2024).

[38] Egeberg and Trondal (n 5).

[39] M Egeberg and J Trondal, "Political Leadership and Bureaucratic Autonomy: Effects of Agencification" (2009) 22 Governance 673.

### *VI.2. Integrating the Forum and the Panel into a single body*

The first point concerns the institutional framework of the advisory bodies and stakeholder representation in them. As anticipated, there is potential for consolidating the Panel and the Forum into a singular entity. Integrating them would reduce duplications and bolster the deliberation process before reaching a decision. A combined entity would merge the diverse knowledge bases of civil society, the business sector, and the academic community, promoting inclusive and reflective discussions of the needs identified by the Commission and Member States. A unified entity combining the Advisory Forum's extensive stakeholder engagement with the Scientific Panel's specialised, independent expertise could significantly improve the quality of advice to the Board, the Office, and other EU institutions or agencies. The unified entity would ensure that the guidance reflects both the technical complexities and societal implications of AI and challenges the belief that GPAI necessitates fundamentally different knowledge from other AI systems. Subcommittees or working groups could help avoid the risk of this unified body becoming overburdened or diluting specific expertise within a larger group.

Should merging the Advisory Forum and the Scientific Panel prove infeasible, an alternative solution could be to better coordinate their operations, for example, through clear separations of scopes, roles, and tasks, but unified reporting. While not as ideal as a complete merger, this approach could streamline the reporting process by creating a common framework for both groups to communicate their findings and recommendations. This might entail producing a joint annual report consolidating contributions from the Forum and the Panel, thereby cutting administrative overlap and ensuring a more unified advisory voice to the Commission, the Board, and the Member States.

Merging or enhancing coordination between the Advisory Forum and Scientific Panel, complemented by creating subcommittees, favors robust governance of the AIA by streamlining advisory roles for agility and innovation, also in response to disruptive technological changes.

---

[40] M Scholten and M van Rijsbergen, "The Limits of Agencification in the European Union" (2014) 15 German Law Journal 1223; M Chamon, "Setting the Scene: EU Agencies, Agencification, and the EU Administration" in M Chamon (ed), *EU Agencies: Legal and Political Limits to the Transformation of the EU Administration* (Oxford University Press 2016) <https://doi.org/10.1093/acprof:oso/9780198784487.003.0002> accessed 29 January 2024; A Koene and others, "A Governance Framework for Algorithmic Accountability and Transparency" <https://nottingham-repository.worktribe.com/index.php/output/3979928/a-governance-framework-for-algorithmic-accountability-and-transparency> (accessed 3 February 2022).

### VI.3. Coordinating overlapping EU entities: the case for an AI Coordination Hub

As AI technologies proliferate across the EU, collaboration among various regulatory entities becomes increasingly critical, especially when introducing new AI applications intersects with conflicting interests. A case in point is the independent decision by Italy's data protection authority, 'Garante per la privacy', to suspend ChatGPT, a move not mirrored by other data protection entities within the EU.[41] The scope for such overlaps is not limited to national data protection authorities but extends to other entities, such as decentralised agencies, including the European Data Protection Board (EDPB), the European Union Agency for Cybersecurity (ENISA), the Fundamental Rights Agency (FRA), the European Medicines Agency (EMA), the European Banking Authority (EBA), and the European Union Intellectual Property Office (EUIPO). The likelihood of overlaps and interferences with the constellation of bodies now introduced by the AIA – e.g., the Office, the Board, the Forum, etc. – is high.

In light of these challenges, it becomes crucial to incorporate efficient coordination mechanisms within the EU's legislative framework. Enhancing the functionality of the existing EU Agency Network[42], to foster a collaborative environment and act as a unified point of communication for all EU agencies and Joint Undertakings (JUs) on multifaceted issues, would mark a significant advancement. However, establishing a centralised platform, the European Union Artificial Intelligence Coordination Hub (EU AICH), emerges as a compelling alternative. This hub would convene all pertinent bodies involved in AI regulation and oversight, facilitating collective decision-making. Establishing such a hub promises to elevate significantly the uniformity of AIA enforcement, improve operational efficiency, and reduce inconsistencies in treating similar matters.

### VI.4. Control of AI misuse at the EU level

The AI Board's lack of authority to revise or address decisions made by national authorities, in contrast to the European Data Protection Board's role under GDPR, creates a significant gap in ensuring consistent AI regulation across the EU. This deficiency could result in divergent interpretations and applications of the AIA, similar to the challenges observed with the GDPR, particularly in cases such as the Irish Data Protection Commission's oversight of major tech firms. Such inconsistencies are concerning, especially regarding the AIA's restrictions on surveillance tools, including facial recognition technologies. Without the ability to correct or harmonise national decisions, there's a heightened risk that AI could be misused in some Member States, potentially facilitating the establishment of illiberal surveillance regimes, and stifling legitimate dissent. This scenario underscores the need for a mechanism within the AI Board to ensure uniform law enforcement and prevent AI's abusive applications, especially in sensitive areas like biometric surveillance.

### VI.5. Learning mechanisms

Given their capacity for more rapid development and adjustment, the agility of non-legislative acts presents an opportunity for responsive governance in AI. However, the agility of the regulatory framework must be matched by the regulatory bodies' adaptability. Inter- and intra-agency learning and collaboration mechanisms are essential for addressing AI's multifaceted technical and social challenges.[43] This approach should facilitate continuous improvement and adaptation of regulatory practices to ensure they remain effective in guiding and governing AI technologies' legal and safe development and deployment. To this end, specific ex-ante and ex-post review obligations of the Office's and Board's actions and recommendations could be introduced. More importantly, a dedicated unit, for example, within the AI Office, should be tasked with identifying best and worst practices *across all* involved entities (from the Office to the Forum). Liaising with Member State competence centers, such a unit could become a hub for institutional and individual learning and refinement of AI, within and beyond the AIA framework.

---

[41] Instead, following a request from Noyb, the European Center for Digital Rights, the Austrian Data Protection Agency is set to examine GDPR compliance issues related to ChatGPT, focusing particularly on its tendency to generate inaccurate information. Noyb, "ChatGPT provides false information about people, and OpenAI can't correct it", Noyb blog (29 April 2024), https://noyb.eu/en/chatgpt-provides-false-information-about-people-and-openai-cant-correct-it.

[42] https://agencies-network.europa.eu/index_en.

### *VI.6. (Near) future challenges*

In this paper, we have primarily focused on the structural design of the AIA's institutional framework. However, in the near future, additional challenges will likely arise concerning the interplay between the AIA and other national and international AI regulations. At the EU level, effective enforcement of the AIA will require careful coordination, particularly in light of potential overlaps with implementing other frameworks, such as the GDPR, DMA, and DSA, which are frequently overseen by specialised regulatory authorities.

Internationally, the need for alignment is equally pressing. For instance, how will the AIA's implementation bodies coordinate with those established under US or Chinese legislation? China has enacted several targeted AI regulations, including those explicitly addressing Generative AI.[44] By contrast, the United States currently lacks a formal AI governance model, as the existing legal framework (Biden's Executive Order) has limited scope and enforcement mechanisms.[45] However, it is expected that more stringent federal (e.g., on deep fakes) and state-level legislation (e.g., the California AI Bill) will follow, likely leading to the creation of new governance structures and implementation authorities. Diverging state-level rules may, in fact, result in a problematic patchwork even within the US, making alignment with the AIA even trickier.

Given the United States' dominance of the AI market, the actions of these US authorities will inevitably influence the operations of their EU counterparts. Therefore, to foster constructive collaboration and avoid uncritical emulation, different nation-states and supranational bodies must establish shared objectives and encourage active participation in international forums regarding AI governance, such as the OECD AI Policy Observatory. Such collaboration will help ensure a consistent and effective approach to AI regulation across jurisdictions.

Additionally, the emergence of transnational and international legal documents on AI, such as the Council of Europe's Framework Convention on Artificial Intelligence, raises further considerations. The Convention, drafted by the forty-six Member States of the Council of Europe, along with observer states like the United States, Canada and Japan will be the first-ever international legally binding treaty in this field. It will be crucial to establish a mechanism for interaction between the implementation structures of the AIA and those of the Convention, which includes a follow-up mechanism, the Conference of the Parties. This body, composed of official representatives of the Convention's Parties, periodically monitors the implementation of the Convention's provisions (Article 23 of the

---

[43] Dimitropoulos and Hacker (n 10).

[44] See, e.g., R Creemers, "The Regulation of Generative AI in China, forthcoming" in P Hacker, A Engel, S Hammer, and B Mittelstadt (eds), *The Oxford Handbook of the Foundation and Regulation of Generative AI* (Oxford University Press forthcoming 2024).

[45] M Wörsdörfer, "Biden's Executive Order on AI and the E.U.'s AI Act: A Comparative Computer-Ethical Analysis" (2024) 37 Philosophy & Technology 74.

Convention) and has the authority to resolve disputes between Parties regarding the interpretation or application of the treaty (Article 29 of the Convention).

## VII. Conclusions

An intricate, yet solid foundation for AI governance has been introduced in the AIA. However, this article calls for a forward-looking perspective on AI governance, stressing the importance of anticipatory regulation and the adaptive capabilities of governance structures to keep pace with technological advancements. The article makes five key proposals. First, it suggests establishing the AI Office as a decentralised agency like EFSA or EMA to enhance its autonomy and reduce potential influences from political agendas at the Commission level. Second, there is potential for consolidating the AI Office's advisory bodies – the Advisory Forum and the Scientific Panel – into a single entity to streamline decision-making and improve the quality of advice; this body would reflect both technical and societal implications of AI. Third, the article discusses the need for more coherent decision-making and cooperation among the various EU bodies involved in AI oversight, which may have overlapped or conflicting jurisdictions. This need could be addressed by strengthening the existing EU Agency Network or creating an EU AI Coordination Hub. Fourth, the lack of authority for the AI Board to revise national decisions could lead to inconsistent application of AI regulations across Member States, like issues observed with GDPR enforcement. Fifth, to ensure responsive and effective governance of AI technologies, it proposes introducing mechanisms for continuous learning and adaptation within the regulatory framework, including a dedicated unit within the AI Office to identify and share best (and worst) practices. The article also calls for simplifying regulatory frameworks to aid compliance, especially for SMEs, and underscores the importance of agile regulatory practices capable of adapting to the rapidly evolving AI landscape, ensuring continuous improvement and effective governance.

Looking ahead, the outlook for the governance of AI in the EU remains promising and challenging. As AI technologies continue to evolve rapidly, the governance structures established by the AIA must remain flexible and adaptive, in short robust, to address new developments and unforeseen risks. Ongoing research, stakeholder engagement, and international cooperation will be essential in refining and updating the regulatory framework. The future of AI governance will likely involve a dynamic balance between providing legal certainty for AI developers and deployers while keeping some terms and concepts strategically vague to cover forthcoming AI advancements. The multilevel structure discussed, ranging from principles in the AIA to rules in delegated and implementing acts, technical standards, and extensive guidance, may combine such safe harbors with open-textured terminology.

**Competing interests.** The author(s) declare none.